\documentclass{article}
\usepackage{spconf,amsmath,epsfig}
\usepackage{amssymb,color,array}
\usepackage{times,epstopdf,textcomp,mathrsfs}
\begin{document}
\ninept

\title{Low-Complexity Robust Adaptive Beamforming Algorithms Based on Shrinkage for Mismatch Estimation }

\name{Hang Ruan * ~and Rodrigo C. de Lamare*$^\#$ \vspace{-0.6em} }

\address{ *Department of Electronics, The University of York, England, YO10 5BB\\
{ $^\#$CETUC, Pontifical Catholic University of Rio de Janeiro, Brazil}\\
 Emails: hang.ruan@york.ac.uk,  delamare@cetuc.puc-rio.br \vspace{-0.85em}
 \sthanks{This work was supported in part by The
University of York}}

\maketitle

\begin{abstract}
In this paper, we propose low-complexity robust adaptive beamforming
(RAB) techniques that based on shrinkage methods. The only prior
knowledge required by the proposed algorithms are the angular sector
in which the actual steering vector is located and the antenna array
geometry. We firstly present a Low-Complexity Shrinkage-Based
Mismatch Estimation (LOCSME) algorithm to estimate the desired
signal steering vector mismatch, in which the
interference-plus-noise covariance (INC) matrix is estimated with
Oracle Approximating Shrinkage (OAS) method and the weights
are computed with matrix inversions. We then develop low-cost
stochastic gradient (SG) recursions to estimate the INC matrix and update
the beamforming weights, resulting in the proposed LOCSME-SG
algorithm. Simulation results show that both LOCSME and
LOCSME-SG achieve very good output signal-to-interference-plus-noise
ratio (SINR) compared to previously reported adaptive RAB
algorithms.
\end{abstract}


\section{Introduction}

Several important applications of adaptive beamforming like wireless
communications, radar and sonar, microphone array processing have
been intensively studied in the past years. However, under certain
circumstances, adaptive beamformers may suffer performance
degradation due to short data records, the presence of the desired
signal in the training data, or imprecise knowledge of the desired
signal steering vector. In order to address these problems, robust
adaptive beamforming (RAB) techniques have been developed in recent
years \cite{r1}-\cite{r11}. From a design principle point of view,
the generalized sidelobe canceller, worst-case optimization
\cite{r3}, diagonal loading \cite{r4,r5}, eigenspace projection and
steering vector estimation with presumed prior knowledge
\cite{r7,r8} have been investigated. However, RAB designs based on
these principles have some drawbacks such as their ad hoc nature,
high probability of subspace swap at low SNR and high computational
cost \cite{r7}.

Recent works have focused on design approaches that combine different
principles together to improve RAB performance. Methods which jointly estimate
the mismatched steering vector using Sequential Quadratic Program (SQP)
\cite{r8} and the interference-plus-noise covariance (INC) matrix using a
shrinkage method \cite{r10} have been reported. Another similar approach which
jointly estimates the steering vector using an SQP and the INC matrix using a
covariance reconstruction method \cite{r11}, presents outstanding performance
compared to other RAB techniques. However, their main disadvantage is the high
computational cost associated with the optimization algorithms \cite{r10,r11}
and the matrix reconstruction process \cite{r11},\cite{cisco}-\cite{viswanath}.

This paper proposes adaptive RAB algorithms with low complexity,
which require very little in terms of prior information and shows a
better performance than previously reported algorithms. Firstly, the
steering vector of the desired signal is estimated using a
Low-Complexity Shrinkage-Based Mismatch Estimation (LOCSME)
algorithm. An extension of the Oracle Approximating Shrinkage (OAS)
method \cite{r12} is employed to perform shrinkage estimation of the
cross-correlation vector between the sensor array received data and
the beamformer output. The mismatched steering vector is then
efficiently estimated without any costly optimization procedure in a
low-complexity sense. Secondly, we estimate the desired signal power
using the desired signal steering vector and the input data. We also
develop a stochastic gradient (SG) version of LOCSME, denoted
LOCSME-SG, which does not require matrix inversions or costly
recursions. In particular, in LOCSME-SG the INC matrix from the
input data is estimated using a Knowledge-Aided (KA) shrinkage
\cite{r15} approach along with the computation of the beamforming
weights based on the estimated steering vector through SG
recursions. The proposed LOCSME and LOCSME-SG algorithms circumvent
the use of direction finding techniques for the interferers when
obtaining the INC matrix and only require the angular sector in
which the desired signal steering vector lies as prior knowledge.

This paper is structured as follows. The system model and problem
statement are described in Section ${\bf 2}$. The proposed LOCSME
and LOCSME-SG algorithms are introduced in Sections ${\bf 3}$ and
${\bf 4}$, respectively. Section ${\bf 5}$ presents and discusses
the simulation results. Section ${\bf 6}$ gives the conclusion.

\section{System Model and Problem Statement}

Consider a linear antenna array of $M$ sensors and $K$ narrowband signals. The data received at the $i$th snapshot can be modeled as
\begin{equation}
{\bf x}(i)={\bf A}({\boldsymbol \theta}){\bf s}(i)+{\bf n}(i),
\end{equation}
where ${\bf s}(i) \in {\mathbb C}^{K \times 1}$ are uncorrelated
source signals,
${\boldsymbol\theta}=[{\theta}_1,\dotsb,{\theta}_K]^T \in {\mathbb
R}^K$ is a vector containing the directions of arrival (DoAs), ${\bf
A}({\boldsymbol \theta})=[{\bf a}({\theta}_1 )+{\bf e}, \dotsb, {\bf
a}({\theta}_K)] \in {\mathbb C}^{M \times K}$ is the matrix which
contains the steering vector for each DoA and ${\bf e}$ is the
steering vector mismatch of the desired signal, ${\bf n}(i) \in
{\mathbb C}^{M \times 1}$ is assumed to be complex Gaussian noise
with zero mean and variance ${\sigma}^2_n$. The beamformer output is
given by
\begin{equation}
y(i)={\bf w}^H{\bf x}(i),
\end{equation}
where ${\bf w}=[w_1,\dotsb,w_M]^T \in {\mathbb C}^{M\times1}$ is the
beamformer weight vector, where $({\cdot})^H$ denotes the Hermitian
transpose. The optimum beamformer is computed by maximizing the
signal-to-interference-plus-noise ratio (SINR) given by
\begin{equation}
SINR=\frac{{\sigma}^2_1{\lvert{\bf w}^H{\bf a}\rvert}^2}{{\bf w}^H{\bf R}_{i+n}{\bf w}}.
\end{equation}
where ${\sigma}^2_1$ is the desired signal power, ${\bf R}_{i+n}$ is the INC matrix and assume the steering vector ${\bf a}$ is known precisely (${\bf a}={\bf a}({\theta}_1 )$), then problem (3) can be transformed into an optimization problem as
\begin{equation}
\begin{aligned}
& \underset{\bf w} {\text{minimize}}
&& {\bf w}^H{\bf R}_{i+n}{\bf w} \\
& \text{subject to} && {\bf w}^H{\bf a}=1,
\end{aligned}
\end{equation}
which is known as the MVDR beamformer or Capon beamformer \cite{r1}. The optimum weight vector is given by ${\bf w}_{opt}=\frac{{{\bf R}^{-1}_{i+n}}{\bf a}}{{\bf a}^H{{\bf R}^{-1}_{i+n}}{\bf a}}.$ Since ${\bf R}_{i+n}$ is usually unknown in practice, it can be estimated by the sample covariance matrix (SCM) of the received data as
\begin{equation}
\hat{\bf R}(i)=\frac{1}{i}\sum\limits_{k=1}^i{\bf x}(k){{\bf x}^H}(k),
\end{equation}
which results in the Sample Matrix Inversion (SMI) beamformer ${\bf
w}_{SMI}=\frac{\hat{\bf R}^{-1}{\bf a}}{{\bf a}^H\hat{\bf
R}^{-1}{\bf a}}$. However, the SMI beamformer requires a large
number of snapshots to converge and is sensitive to steering vector
mismatches \cite{r10,r11}. The problem we are interested in solving
is how to design low-complexity robust beamforming algorithms that
can preserve the SINR performance in the presence of uncertainties
in the steering vector of a desired signal.

\section{Proposed LOCSME Algorithm}

In this section, the proposed LOCSME algorithm is introduced. The
basic idea of LOCSME is to obtain a precise estimate of the desired
signal steering vector and afterwards use it to estimate the desired
signal power and to derive the recursion for the weight vector. The
estimation of the steering vector is described as the projection
onto a predefined subspace matrix of an iteratively
shrinkage-estimated cross-correlation vector between the beamformer
output and the array observation data. To obtain the INC matrix, in
LOCSME we use OAS method to shrink the SCM in order to estimate the
INC matrix.

\subsection{{Steering Vector Estimation}}

The cross-correlation between the array observation data and the
beamformer output can be expressed as ${\bf d}=E\lbrace{\bf
x}y^*\rbrace$. With assumptions that ${\lvert{{\bf a}_m{\bf
w}}\rvert}\ll{\lvert{{\bf a}_1{\bf w}}\rvert}$ for $m=2,\dotsb,K$
and that signal sources and that the system noise have zero mean
while the desired signal is independent from the interferers and the
noise, ${\bf d}$ can be rewritten as ${\bf
d}=E\lbrace{{{{\sigma}_1}^2{\bf a}_1^H{\bf w}{\bf a}_1}+{\bf n}{\bf
n}^H{\bf w}}\rbrace$. By projecting ${\bf d}$ onto a predefined
subspace \cite{r9} which collects all possible information from the
desired signal, the unwanted part of ${\bf d}$ can be eliminated.
The prior knowledge amounts to providing an angular sector in which
the desired signal is located, say
$[{\theta}_1-{\theta}_e,{\theta}_1+{\theta}_e]$. The subspace
projection matrix ${\bf P}$ is given by
\begin{equation}
{\bf P}=[{\bf c}_1,{\bf c}_2,\dotsb,{\bf c}_p][{\bf c}_1,{\bf c}_2,\dotsb,{\bf c}_p]^H,
\end{equation}
where ${\bf c}_1,\dotsb,{\bf c}_p$ are the $p$ principal eigenvectors vectors of the matrix ${\bf C}$, which is defined by \cite{r8}
\begin{equation}
{\bf C}= \int\limits_{{\theta}_1-{\theta}_e}^{{\theta}_1+{\theta}_e}{\bf a}({\theta}){\bf a}^H({\theta})d{\theta}.
\end{equation}
We then employ the OAS shrinkage technique in order to achieve a
more accurate estimation of ${\bf d}$, so that it can help us to
obtain a better estimate of the steering vector. Let us define
\begin{equation}
\hat{\bf F}=\hat{\nu} {\bf I},
\end{equation}
where $\hat{\nu} = {{\rm tr}(\hat{\bf S})}/M$ and $\hat{\bf S} =
{\rm diag}({\bf x}y^*)$. By shrinking $\hat{\bf S}$ towards
$\hat{\bf F}$ \cite{r12} and subsequently using it in a vector
shrinkage form, taking into account the snapshot index, the result
gives
\begin{equation}
\hat{\bf d}(i) = \hat{\rho}(i){\rm diag}(\hat{\bf F}(i)) + (1-\hat{\rho}(i)){\rm diag}(\hat{\bf S}(i)),
\end{equation}
which is parameterized by the shrinkage coefficient $\hat{\rho}(i)$.
If we define $\hat{\bf D}={\rm diag}(\hat{\bf d})$, then the goal is
to find the optimal value of $\hat{\rho}(i+1)$ that minimizes the
mean square error (MSE) of $E[{\lVert{\hat{\bf D}(i+1)-\hat{\bf
F}(i)}\rVert}^2]$ in the $i$th snapshot, which leads to
\begin{equation}
\hat{\rho}(i+1) = \frac{(1-\frac{2}{M}){\rm tr}(\hat{\bf D}(i)\hat{\bf S}^*(i)) + {\rm tr}(\hat{\bf D}(i)){\rm tr}(\hat{\bf D}^*(i))}{(i+1-\frac{2}{M}){\rm tr}(\hat{\bf D}(i)\hat{\bf S}^*(i))+(1-\frac{i}{M}){\rm tr}(\hat{\bf D}(i)){\rm tr}(\hat{\bf D}^*(i))},
\end{equation}
where the derivation is shown in the Appendix and $\hat{\bf
S}(i)={\rm diag}(\hat{\bf l}(i))$, where $\hat{\bf
l}(i)=\frac{1}{i}\sum\limits_{k=1}^i{\bf x}(k)y^*(k)$, is the sample
correlation vector (SCV). Alternatively equation (10) can be
re-expressed in vector multiplication form and leads to the
following
\begin{equation}
\hat{\rho}(i+1) = \frac{(1-\frac{2}{M})\hat{\bf d}^H(i)\hat{\bf l}(i) + {\rm tr}(\hat{\bf D}(i)){\rm tr}(\hat{\bf
D}^*(i))}{(i+1-\frac{2}{M})\hat{\bf d}^H(i)\hat{\bf l}(i)+(1-\frac{i}{M}){\rm tr}(\hat{\bf D}(i)){\rm tr}(\hat{\bf
D}^*(i))},
\end{equation}
As long as the initial value of $\hat{\rho}(0)$ is between $0$ and $1$, the iterative process in (9) and (11) is guaranteed to converge \cite{r12}. Once the correlation vector $\hat{\bf d}$ is obtained, the steering vector is estimated by
\begin{equation}
{\hat{\bf a}_1}(i)=\frac{{\bf P}\hat{\bf d}(i)}{{\lVert{{\bf P}\hat{\bf d}(i)}\rVert}_2}.
\end{equation}

\subsection{{Desired Signal Power Estimation}}

This subsection will exploit a novel method to estimate the desired
signal power ${\sigma}^2_1$. This can be accomplished by directly
using the desired signal steering vector. Let us rewrite the
received data as
\begin{equation}
{\bf x}(i)=\hat{\bf a}_1(i){s_1}+\sum\limits_{k=2}^K{\bf a}_k{s_k}+{\bf n}(i).
\end{equation}
Pre-multiplying the above equation by $\hat{\bf a}_1^H(i)$ and
assuming $\hat{\bf a}_1(i)$ is uncorrelated with the interferers, we
obtain
\begin{equation}
\hat{\bf a}_1^H(i){\bf x}(i)=\hat{\bf a}_1^H(i)\hat{\bf a}_1(i){s_1}+\hat{\bf a}_1^H(i){\bf n}(i).
\end{equation}
Taking the expectation $|\hat{\bf a}_1^H(i){\bf x}(i)|^2$, we obtain
\begin{multline}
|\hat{\bf a}_1^H(i){\bf x}(i)|^2=E[(\hat{\bf a}_1^H(i)\hat{\bf a}_1(i){s_1}+\hat{\bf a}_1^H(i){\bf n}(i))^* \\ (\hat{\bf a}_1^H(i)\hat{\bf a}_1(i){s_1}+\hat{\bf a}_1^H(i){\bf n}(i))].
\end{multline}
If the noise is statistically independent from the desired signal, then we have
\begin{equation}
|\hat{\bf a}_1^H(i){\bf x}(i)|^2=|\hat{\bf a}_1^H(i)\hat{\bf a}_1(i)|^2|s_1|^2+\hat{\bf a}_1^H(i)E[{\bf n}(i){\bf n}^H(i)]\hat{\bf a}_1(i),
\end{equation}
where $E[{\bf n}(i){\bf n}^H(i)]$ represents the noise covariance
matrix ${\bf R}_n$ which can be replaced by ${\sigma}^2_n{\bf I}_M$,
where ${\sigma}^2_n$ is assumed known here for convenience,
otherwise it can be easily estimated by a specific estimation
method. Replacing the desired signal power $|s_1|^2$ by its estimate
$\hat{\sigma}^2_1(i)$, the desired signal power estimate is computed
as
\begin{equation}
\hat{\sigma}^2_1(i)=\frac{|{{\hat{\bf a}_1}^H}(i){\bf x}(i)|^2-{{\hat{\bf a}_1}^H}(i){\hat{\bf a}_1}(i){\sigma}^2_n}{|{{\hat{\bf a}_1}^H}(i){\hat{\bf a}_1}(i)|^2}.
\end{equation}
Equation (17) has a low complexity ($\mathcal{O}(M)$) and can be
directly implemented if the desired signal steering vector is well
estimated and the noise level is known.

\subsection{{Estimation of the INC matrix}}

In this subsection we describe a method to estimate the INC matrix
that is based on the OAS method \cite{r12} and used in LOCSME. In
the OAS estimation of LOCSME, we need the SCM in (5) as a
preliminary estimate for the INC matrix. Then we define $\hat{\bf
F}_0={\hat{\nu}_0}{\bf I}$, where $\hat{\nu}_0={{\rm tr}(\hat{\bf
R})}/M$. By minimizing the MSE described by $E[{\lVert{{\tilde{\bf
R}(i)}-{\hat{\bf F}_0}(i-1)}\rVert}^2]$ \cite{r12}, the following
recursion is employed:
\begin{equation}
\tilde{\bf R}(i)={\hat{\rho}_0}(i){\hat{\bf F}_0}(i)+(1-{\hat{\rho}_0}(i))\hat{\bf R}(i),
\end{equation}
\begin{equation}
{\hat{\rho}_0}(i+1)=\frac{(1-{\frac{2}{M}}){\rm tr}({\tilde{\bf
R}(i)}{\hat{\bf R}(i)})+{\rm tr}^2(\tilde{\bf
R}(i))}{(i+1-{\frac{2}{M}}){\rm tr}({\tilde{\bf R}(i)}{\hat{\bf
R}(i)})+(1-{\frac{i}{M}}){\rm tr}^2(\tilde{\bf R}(i))},
\end{equation}
where ${\hat{\rho}_0}(0)$ must be initialized between $0$ and $1$ to
guarantee convergence \cite{r12}. To exclude the information of the
desired signal from the covariance matrix of the array observation
data, a simple subtraction is considered:
\begin{equation}
{\tilde{\bf R}_{i+n}}(i)=\tilde{\bf R}(i)-\hat{\sigma}^2_1(i){\hat{\bf
a}_1}(i){{\hat{\bf a}_1}^H}(i).
\end{equation}

\subsection{{Computation of Beamforming Weights}}

For the proposed LOCSME algorithm the beamforming weights are computed directly by
\begin{equation}
\hat{\bf w}(i)=\frac{{\tilde{\bf R}^{-1}_{i+n}}(i)\hat{\bf a}_1(i)}{{\hat{\bf a}_1}^H(i){\tilde{\bf R}^{-1}_{i+n}}(i)\hat{\bf a}_1(i)},
\end{equation}
which has a computational costly matrix inversion ${\tilde{\bf
R}^{-1}_{i+n}}(i)$. to reproduce the proposed LOCSME algorithm,
whose complexity is ${\mathcal{O}}(M^{3})$, equations (9),(11),(12)
and (17)-(21) are required. In comparison to previously reported RAB
algorithms in \cite{r7,r8,r10,r11} with costly online optimization
procedures and complexity ${\mathcal{O}}(M^{3})$ or higher, LOCSME
requires a similar or lower cost.

\section{Proposed LOCSME-SG Algorithm}

In this section, the proposed LOCSME-SG algorithm is detailed. The
aim is to devise a low-complexity alternative to LOCSME that is
suitable for time-varying scenarios and implementation purposes.
LOCSME-SG employs identical recursions to LOCSME to estimate the
steering vector and the desired signal power, whereas the estimation
of the INC matrix and the beamforming weights is different. In
particular, LOCSME-SG employs a Modified Array Observation (MAO)
vector to compute a preliminary estimate of the INC matrix, followed
by a refined estimate with a low-cost shrinkage method.

\subsection{{Estimation of the INC matrix}}

In this subsection we present an extension of the KA shrinkage
method \cite{r15} to estimate the INC matrix, which has much lower
complexity than the one used in LOCSME. In LOCSME-SG, with the
estimate of the desired signal power we subtract unwanted
information of the interferences out from the array received data in
a low complexity vector form to obtain the MAO vector. Consider a
simple substraction step as
\begin{equation}
{\bf x}_{i+n}(i)={\bf x}(i)-\hat{\sigma}_1(i)\hat{\bf a}_1(i).
\end{equation}
Then the INC matrix can be estimated by
\begin{equation}
\hat{\bf R}_{i+n}(i)={\bf x}_{i+n}(i){\bf x}_{i+n}^H(i).
\end{equation}
Now, we employ the idea of KA shrinkage method \cite{r15} to help
with our INC estimation. By applying a linear shrinkage model for
the INC matrix, we have
\begin{equation}
\tilde{\bf R}_{i+n}(i)=\eta(i){\bf R}_0+(1-\eta(i))\hat{\bf
R}_{i+n}(i),
\end{equation}
where ${\bf R}_0$ is an initial guess for the INC matrix, $\eta(i)$
is the shrinkage parameter and $\eta(i) \in (0,1)$. Here the
shrinkage parameter is expected to be adaptively estimated.
Employing an idea of adaptive filtering \cite{r15}, it is possible
to set the overall filter output $y_f(i)$ equal to $[\tilde{\bf
R}_{i+n}(i)\hat{\bf a}_1(i)]^H{\bf x}(i)$ which is the linear
combination of the outputs from two filter elements which are
$y_{0f}(i)=[{\bf R}_0\hat{\bf a}_1(i)]^H{\bf x}(i)$ and ${\hat
y_f}(i)=[\hat{\bf R}_{i+n}(i)\hat{\bf a}_1(i)]^H{\bf x}(i)$, which
leads to
\begin{equation}
y_f(i)=\eta(i)y_{0f}(i)+(1-\eta(i)){\hat y_f}(i).
\end{equation}
To restrict the value of $\eta(i)$ equal to either $0$ nor $1$, a
sigmoidal function is employed:
\begin{equation}
\eta(i)={\rm sgm}[\epsilon(i)]=\frac{1}{1+e^{-\epsilon(i)}},
\end{equation}
where $\epsilon(i)$ is updated as
\begin{multline}
\epsilon(i+1)=\epsilon(i)-\frac{\mu_{\epsilon}}{(\sigma_{\epsilon}+q(i))}(\eta(i)|y_{0f}(i)-{\hat
y_f}(i)|^2 \\ +\mathcal{R}\lbrace{(y_{0f}(i)-{\hat y_f}(i)){\hat
y_f}^*(i)}\rbrace)\eta(i)(1-\eta(i)),
\end{multline}
where $\mu_{\epsilon}$ is the step size while $\sigma_{\epsilon}$ is
a small positive constant, $q(i)$ is updated as
\begin{equation}
q(i+1)={\lambda}_q(i)(1-{\lambda}_q)|y_{0f}(i)-{\hat y_f}(i)|^2,
\end{equation}
where ${\lambda}_q$ is a forgetting factor \cite{r15}.

\subsection{Computation of Beamforming Weights}

For the proposed LOCSME-SG algorithm, we resort to an SG adaptive
algorithm to reduce the complexity required by the matrix inversion.
The optimization problem (4) can be re-expressed as
\begin{equation}
\begin{aligned}
& \underset{{\bf w}(i)} {\text{minimize}}
&& {\bf w}^H(i)({\bf x}(i){\bf x}^H(i)-\hat{\sigma}^2_1(i){\hat{\bf a}_1}(i){\hat{\bf a}^H_1}(i)){\bf w}(i) \\
& \text{subject to} && {\bf w}^H(i)\hat{\bf a}_1(i)=1.
\end{aligned}
\end{equation}
In order to compute the beamforming weights, we employ an SG
recursion as given by
\begin{equation}
{\bf w}(i+1)={\bf w}(i)-\mu\frac{\partial\mathcal{L}}{\partial{\bf w}(i)},
\end{equation}
where ${\mathcal{L}}={\bf w}^H(i)({\bf x}(i){\bf
x}^H(i)-\hat{\sigma}^2_1(i){\hat{\bf a}_1}(i){\hat{\bf
a}^H_1}(i)){\bf w}(i)+\lambda({\bf w}^H(i){\hat{\bf a}_1}(i)-1)$. By
substituting $\mathcal{L}$ into the SG equation (30) and letting
${\bf w}^H(i+1)\hat{\bf a}_1(i+1)=1$, $\lambda$ is obtained as
\begin{equation}
\lambda=\frac{y(i){\bf x}^H(i){\hat{\bf a}^H_1}(i)-\hat{\sigma}^2_1(i){\hat{\bf a}^H_1}(i){\hat{\bf a}_1}(i)}{{\hat{\bf a}^H_1}(i){\hat{\bf a}_1}(i)}.
\end{equation}
By substituting (31) back into (30) again, the weight update
equation for LOCSME-SG is obtained as
\begin{multline}
{\bf w}(i+1)=({\bf I}-\mu\hat{\sigma}^2_1(i){\hat{\bf
a}_1}(i){\hat{\bf a}^H_1}(i)){\bf w}(i)-\mu(({\bf I}+\frac{{\hat{\bf
a}_1}(i){\hat{\bf a}^H_1}(i)}{{\hat{\bf a}^H_1}(i){\hat{\bf
a}_1}(i)}) \\ y^*(i){\bf x}(i)-\hat{\sigma}^2_1(i){\hat{\bf
a}_1}(i)).
\end{multline}
The adaptive SG recursion circumvents a matrix inversion when
computing the weights using (21), which is unavoidable in LOCSME.
Therefore, the computational complexity for computing the weights is
reduced from ${\mathcal{O}}(M^3)$ in LOCSME to ${\mathcal{O}}(M^2)$
in LOCSME-SG ($15M^2+25M$). The proposed LOCSME-SG algorithm can be
reproduced by using equations (9),(11),(12),(17),(22)-(28) and (32).
Moreover, compared to existing RAB algorithms
\cite{r4,r7,r8,r9,r10,r11} and LOCSME which have a complexity equal
or higher than ${\mathcal{O}}(M^{3})$, LOCSME-SG has a greatly
reduced cost. Compared with the approach in \cite{r14} which
implements a low-complexity worst-case adaptive algorithm with a
computational complexity of $2M^2+7M$, LOCSME-SG achieves a much
better performance.


\section{Simulations}

A uniform linear array (ULA) of $M=12$ omnidirectional sensors with
a spacing of half wavelength is considered in the simulations. The
desired signal is assumed to arrive at ${\theta}_1=10^\circ$ while
there are other two interferers impinging on the antenna array from
directions ${\theta}_2=50^\circ$ and ${\theta}_3=90^\circ$. The
signal-to-interference ratio (SIR) is fixed at 20dB. We employ 100
repetitions to obtain each point of the curves and allow only one
iteration performed per snapshot. In our algorithm, the angular
sector is chosen as $[{\theta}_1-5^\circ,{\theta}_1+5^\circ]$ and
the number of eigenvectors of the subspace projection matrix $p$ is
selected manually with the help of simulations. The proposed LOCSME
and LOCSME-SG algorithms are compared with previously developed
low-complexity standard SG algorithm and the RAB method which is
based on the worst-case optimization \cite{r14}. We consider both
coherent local scattering and incoherent local scattering scenarios
for the mismatch and look at the beamformer output SINR in terms of
snapshots with a maximum of $i=500$ snapshots observed, or a
variation of input SNR (-10dB to 30dB) as shown in Figs. 1 and 2,
respectively.

\subsection{{Mismatch due to Coherent Local Scattering}}

The steering vector of the desired signal affected by a local scattering effect is modeled as
\begin{equation}
{\bf a}={\bf p}+\sum\limits_{k=1}^4{e^{j{\varphi}_k}}{\bf b}({\theta}_k),
\end{equation}
where ${\bf p}$ corresponds to the direct path while ${\bf b}({\theta}_k)(k=1,2,3,4)$ corresponds to the scattered paths. The angles ${\theta}_k(k=1, 2, 3, 4)$ are randomly and independently drawn in each simulation run from a uniform generator with mean $10^\circ$ and standard deviation $2^\circ$. The angles ${\varphi}_k(k=1, 2, 3, 4)$ are independently and uniformly taken from the interval $[0,2\pi]$ in each simulation run. Notice that ${\theta}_k$ and ${\varphi}_k$ change from trials while remaining constant over snapshots \cite{r3}. We select $\mu=0.2$, ${\mu}_{\epsilon}=1$, ${\sigma}_{\epsilon}=0.001$, ${\lambda}_q=0.99$, ${\bf R}_0=10{\bf I}$. The SINR performance versus snapshots and SNR of all the tested algorithms affected by coherent scattering is illustrated as  in Figs. 1 (a) and 2 (a). LOCSME-SG outperforms the other algorithms and is very close to the standard LOCSME.

\subsection{{Mismatch due to Incoherent Local Scattering}}

In the incoherent local scattering case, the desired signal has a time-varying signature and the steering vector is modeled by
\begin{equation}
{\bf a}(i)=s_0(i){\bf p}+\sum\limits_{k=1}^4{s_k(i)}{\bf b}({\theta}_k),
\end{equation}
where $s_k(i)(k=0, 1, 2, 3, 4)$ are i.i.d zero mean complex Gaussian random variables independently drawn from a random generator. The angles ${\theta}_k(k=0, 1, 2, 3, 4)$ are drawn independently in each simulation run from a uniform generator with mean $10^\circ$ and standard deviation $2^\circ$. This time, $s_k(i)$ changes both from run to run and from snapshot to snapshot. We select $\mu=0.1$, ${\mu}_{\epsilon}=5$, ${\sigma}_{\epsilon}=0.001$, ${\lambda}_q=0.99$, ${\bf R}_0=50{\bf I}$. The SINR performance versus snapshots and SNR is depicted in Figs. 1 (b) and 2 (b). Different from the coherent scattering results, all the algorithms have a certain level of performance degradation due to the effect of incoherent local scattering. However, over a wide range of input SNR, LOCSME-SG is able to outperform the standard SG and the low-complexity worst-case beamformers.

\begin{figure}[!htb]
\begin{center}
\def\epsfsize#1#2{0.95\columnwidth}
\epsfbox{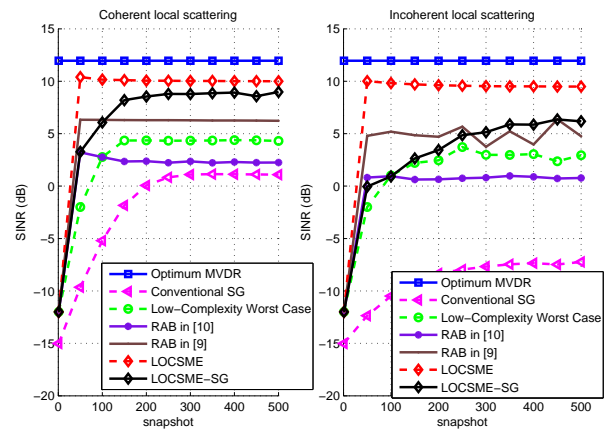} \vspace*{-1.5em} \caption{SINR versus snapshots.} \label{1}
\end{center}
\end{figure}

\begin{figure}[!htb]
\begin{center}
\def\epsfsize#1#2{0.95\columnwidth}
\epsfbox{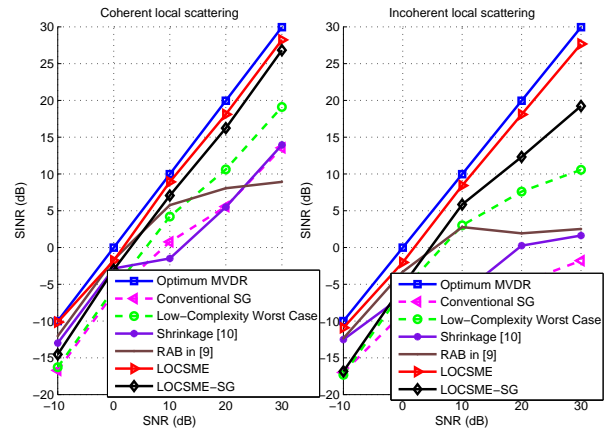} \vspace*{-1.5em} \caption{SINR versus SNR.} \label{2}
\end{center}
\end{figure}

\section{Conclusion}

The proposed LOCSME and LOCSME-SG algorithms only require prior
knowledge of the angular sector of the desired signal and have a low
complexity feature compared to prior art. Compared to the standard
SG beamformer and the low-complexity worst-case approach, LOCSME and
LOCSME-SG have an outstanding performance in both coherent local
scattering and incoherent local scattering cases.

\end{document}